# Enhanced generation of VUV radiation by four-wave mixing in mercury using pulsed laser vaporization

Sébastien Chénais, Sébastien Forget, Laurent Philippet and Marie-Claude Castex

*Laboratoire de Physique des Lasers (UMR 7538 du CNRS), Institut Galilée, Université Paris Nord,*

*99 avenue Jean-Baptiste Clément, 93430 Villetaneuse, France*

Corresponding author: chenais@galilee.univ-paris13.fr ; Tél: (331) 49 40 37 24; fax: (331) 49 40 32 00

**Abstract:** The efficiency of a coherent VUV source at 125 nm, based on 2-photon resonant four-wave mixing in mercury vapor, has been enhanced by up to 2 orders of magnitude. This enhancement was obtained by locally heating a liquid Hg surface with a pulsed excimer laser, resulting in a high density vapor plume in which the nonlinear interaction occurred. Energies up to 5 µJ (1 kW peak power) have been achieved while keeping the overall Hg cell at room temperature, avoiding the use of a complex heat pipe. We have observed a strong saturation of the VUV yield when peak power densities of the fundamental beams exceed the GW/cm$^2$ range, as well as a large intensity-dependant broadening (up to ~30 cm$^{-1}$) of the two-photon resonance. The source has potential applications for high resolution interference lithography and photochemistry.

## 1. Introduction

In the last decades, the generation of coherent tunable VUV light by frequency mixing in gases has received considerable attention both from a fundamental point of view and as an instrument for atomic and molecular spectroscopy [1]. More recently, new applications linked to micro- and nanotechnologies have reinforced the need for practical and "tabletop" sources emitting at the short-wavelength edge of the VUV domain (~110-140 nm): such sources have been thought of as useful tools for, e.g., photolithography [2], processing of wide-band gap materials [3, 4], interference lithography [5], or VUV-induced selective dry etching of semiconductors [6]. The 110-140 nm domain is an interesting region of the spectrum for several reasons: besides the obvious benefit of the diffraction limit compared to existing excimer lasers, the beam is much easier to handle than in the XUV ($\lambda$ < 100 nm) because one can still find transparent optical materials (LiF and MgF$_2$ exhibit cutoff wavelengths around 105 and 120 nm, respectively), and also because a stringent vacuum system is not mandatory (the beam can be manipulated in a O$_2$ and H$_2$O-purged atmosphere, which may be a simple glovebox filled with argon [7]). At last, virtually all materials, including UV transparent dielectrics, interact very strongly with light in this domain [8]. PMMA and PTFE, e.g., exhibit absorption coefficients >10$^5$ cm$^{-1}$ at 125 nm [9] and have accordingly very low ablation thresholds (e.g. ~30 mJ/cm$^2$ at 157 nm for PET [10], presumably less at lower wavelengths.)

Recently, several strategies have been investigated towards the building of coherent sources at such low wavelengths: the growth of nonlinear crystals with transparency range extending far in the VUV has been reported (with reasonable transparency below 150 nm for some borate crystals [11]), as well as a few reports of lasing or stimulated emission in Ar$_2$ excimer media around 126 nm [12, 13]. Sources based on third-order nonlinear frequency conversion in gaseous media remain the most widely used sources, in spite of their general lack of compactness, since they are more versatile, rather efficient (~0.1 % typical) provided that atomic resonances are exploited, and able to operate from the CW [14] to the femtosecond [15] regime, at fixed or tunable wavelengths [16]. If spectroscopy is not the main targeted application, the system does not need to be wavelength-tunable and can be made significantly smaller by using the recent advances in compact diode-pumped solid-state laser and nonlinear optics technologies [17]. Among the nonlinear media frequently used for VUV generation, mercury vapor is especially well-suited for producing tunable radiation around 120 nm by a sum-frequency mixing process, because of its unusually high ionization limit of 84 184 cm$^{-1}$: as a result the high-lying (n>8) $p$ $^1$P$_1$ levels can contribute to a near-three-photon resonant enhancement of the susceptibility. With a tunable laser tuned on an appropriate two-photon resonance, high efficiencies up to ~0.5% [16] have been typically obtained. The highest efficiency ever reported to date is 5% (1.1 mJ @130 nm), obtained by Muller *et al.* [18] from three distinct laser systems, whose wavelengths were chosen to optimize phase matching, in a 1-m long Hg heat pipe. The most recent works on the development of Hg-based sources include the first CW Lyman-alpha source [14, 19] and the achievement of a one-order-of-magnitude enhancement of four-wave mixing using Stark Chirped Rapid Adiabatic Passage [20], with a setup similar to the one presented in this paper. However, all these experiments utilize hot (~200°C) mercury vapor, which restricts their practical use, since complex heat pipes have to be designed to prevent the vapor from coating the windows of the cell (see, e.g., ref [19]). Adding obvious safety and ecological hazards, it is very unlikely that a source relying on hot Hg may find practical applications beyond a precious contribution to spectroscopy.

In this paper, we demonstrate a simple concept to significantly increase the VUV yield while keeping the overall mercury cell at room temperature. The backbone of the source has been described in detail in a previous publication [21], where we reported the generation of VUV light at 125 nm by sum-frequency-mixing in a Hg vapor from one single dye laser. In order to increase the atomic density in the nonlinear interaction zone, instead of globally heating the cell containing the liquid Hg, we use the ability for a laser beam directed onto the Hg surface to create a dense vapor plume in a limited volume (corresponding to the focal region of the mixing fields) and during a very short time (compared to the repetition rate). The average amount of mercury in the cell remains weak and no significant coating of the windows by mercury is observed. It is a similar concept that has been used for high harmonics generation [22] or continuum generation [23] in laser plasmas created by ablation of a solid material. Similarly, four-wave-mixing has been used by Akimov *et al.* [24] to obtain line-by-line imaging of an expanding plasma of optical breakdown. In our case the laser fluence onto the surface has to be sufficiently low (here a few tens of mJ/cm$^2$ corresponding to ~$10^6$ W/cm$^2$ peak power) to guarantee that neutral Hg atoms will outnumber Hg ions and free electrons which do not participate in the resonant mixing process and seriously alter phase matching conditions. This technique seems particularly well suited for mercury, since being a liquid under ambient conditions, the target is not damaged and does not need to be rotated or refreshed between two pulses.

## 2. Experimental

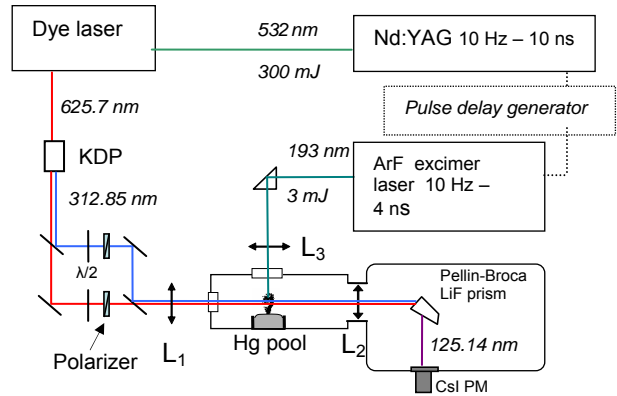

**Figure 1:** experimental setup. See text for details.

The experimental setup appears in Figure 1. A Q-switched frequency-doubled Nd:YAG laser (SAGA from Thales Laser, emitting 300 mJ @532 nm, 10 Hz, 10 ns) was used to pump a pulsed dye laser (TDL 50 from Quantel) working with a 80% DCM /20% Rhodamine 640 dye mixture. Its frequency $\omega_{vis}$ was tuned to 625.700 nm (λ in vacuum, measured with a calibrated wavemeter, Coherent Inc.) so that the two-photon resonance 6s $^1S_0 \rightarrow$ 7s $^1S_0$ was reached with the frequency-doubled output $\omega_{UV} = 2\omega_{vis}$ (λ=312.850 nm). The spectral width of the red beam was measured to be 0.1 cm$^{-1}$ and the pulse duration was 6 ns FWHM. The collinear beams at $\omega_{UV}$ and $\omega_{vis}$ were split thanks to dichroic mirrors and then recombined, which allowed to control the intensities of the two beams independently. The mixing beams were focussed to a 120-μm-in-diameter spot (measured at 1/e$^2$ on the red beam), with a 280-mm focal length achromatic doublet (L$_1$). The harmonic signal at $\omega_{VUV} = 2\omega_{UV} + \omega_{VIS}$ (=125.14 nm) was created in a stainless-steel 30-cm-long cell containing a Hg reservoir. It was primary-pumped to 7.10$^{-3}$ mbar (slightly higher than the Hg vapor pressure at room temperature, 1.8 10$^{-3}$ mbar, that is 4.10$^{13}$ at.cm$^{-3}$) and no buffer gas was added. The VUV beam, after collimation with a LiF lens L$_2$ (f = 170 mm), was separated from the remaining UV and visible beams in an evacuated spectrometer, made of a Pellin-Broca LiF prism, and a CsI solar-blind photomultiplier tube (R2032, Hamamatsu). The spectrometer was calibrated from absolute energy measurements performed at 193 nm with a pyroelectric detector and the spectral sensitivity curve of the detector given by the supplier. We have shown in a previous paper [25] that the polarization of the VUV wave was identical to the polarization of the wave at $\omega_{vis}$. Here a half waveplate inserted in the path of the visible beam enabled the VUV polarization to be horizontal, which maximized the useful measurable signal after the Pellin-Broca prism (measured transmission: 20% at 125 nm).

We chose as a vaporization source an ArF excimer laser (Neweks PSX-100, 3 mJ @193 nm, 4 ns pulse duration) because of the strong absorption coefficient of liquid Hg at this wavelength at normal incidence (60 %). The ArF beam hit the Hg surface via an 85-mm focal-length spherical quartz lens ($L_3$). The focussing of the excimer laser and the height of mixing fields with respect to the pool were adjustable independently. The time delay between the excimer pulse and the fundamental beams was finely tuned with a delay generator (Stanford Research DG535).

### 3. Enhancement of VUV energy

In this section the energies of the waves at $\omega_{UV}$ and $\omega_{vis}$ were set to 3.2 mJ for $\omega_{vis}$ and 1.6 mJ for $\omega_{UV}$ measured inside the Hg cell.

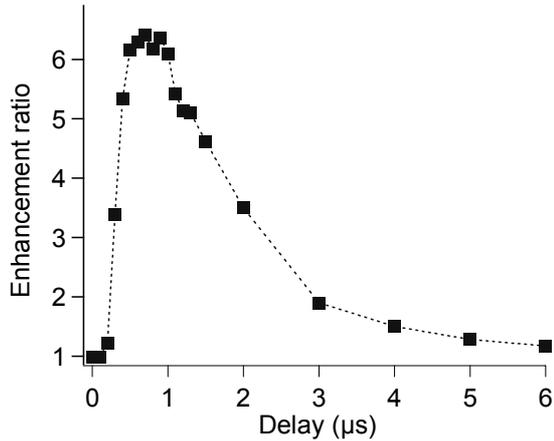

**Figure 2 :** enhancement of VUV yield vs. the delay between the excimer pulse and the VUV pulse.

The VUV enhancement (with respect to the signal obtained in room-temperature mercury without the vaporizing laser) versus time delay appears in Figure 2. The largest enhancement factor (of 6.4) was obtained when the vaporization laser pulse was fired 0.7 µs before the mixing fields, and remained >6 for delays between 0.5 and 1 µs. The measured optimal time delay is consistent with the typical order of magnitude (~µs) observed for the emergence of a high density of neutral atoms with respect to ions and free electrons in a laser-created plasma [26]. We observed that the maximum signal was obtained when UV and red beams grazed the Hg surface; for a higher distance between the surface and the beams, the enhancement followed a rapid decay and reached unity when the focus of mixing fields was 3 mm away from the surface. The enhancement was found to vary with the focussing of the ArF beam: the optimum VUV yield was achieved when the Hg surface was out-of-focus of the quartz lens by 15 mm, the focal point being virtually in the liquid bulk: the excimer pulse (2 mJ, measured in the cell), was then focussed to a 1-mm² square spot. We observed that when the excimer laser was more tightly focussed onto the Hg surface, the VUV enhancement was lower while the plume was emitting a brighter white continuum — clearly visible to the eye — establishing the presence of ionized plasma.

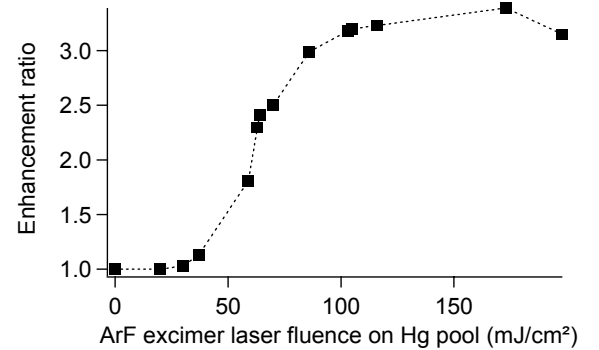

**Figure 3 :** enhancement of VUV yield versus excimer energy. The excimer spot size on the sample is 1 mm².

By varying the energy of the vaporizing laser for a constant spot area (1 mm²), as shown in Figure 3, it is noticeable that a threshold occurs at ~30 mJ/cm² as well as a clear roll-off beyond 100 mJ/cm². The subsequent saturation may be attributed to a change in phase matching conditions associated with a higher degree of ionization of the plasma. A time-resolved analysis of the light emitted by the plume (i.e. a Laser Induced Breakdown Spectroscopy experiment) is under construction in order to gain more insight into the plume dynamics.

A maximum VUV energy per pulse of 5 µJ (~1 kW peak power) inside the cell was obtained when the polarizers and waveplates were removed to maximize the incident energy on the cell. The efficiency, defined following Smith *et al.* [27] as the VUV energy per pulse over the sum of the UV and visible energies, is 0.1 %. This value is of comparable order of magnitude with efficiencies reported in the literature for identical setups [16, 28], i.e. with the same wavelengths for mixing fields, but in which Hg was heated well above room-temperature: for a three-times more powerful UV beam (1.4 MW vs. 0.5 MW peak power in this work), Hilbig *et al.* [16] reported a 0.17% efficiency with a Hg cell heated at 150°C; Mahon and Tomkins [28] reported a 0.6 % efficiency for a mercury cell heated at 190°C, containing 15 torr of helium as a buffer gas, and with an order-of-magnitude higher energy for the visible beam.

## 4. Saturation issues

In this section we study the evolution of the VUV enhancement with the intensity of the mixing fields. The excimer energy is kept fixed to 2 mJ. The intensity of waves at $\omega_{UV}$ and $\omega_{vis}$ were adjustable with the half-waveplate/polarizer assemblies depicted in Figure 1.

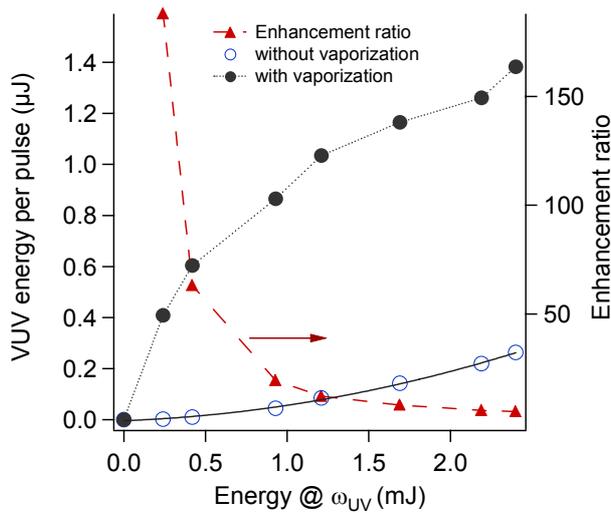

**Figure 4 :** *left axis*: VUV energy per pulse obtained for a fixed energy of red beam (3.4 mJ). *Open circles*: without vaporization; the line is a quadratic fit showing that no saturation occurs. *Full circles*: with vaporization (the line is a guide for the eye). *Right axis*: enhancement ratio (VUV energy with vaporization / energy without vaporization). The excimer energy is 2 mJ (0.2 J/cm² on Hg surface).

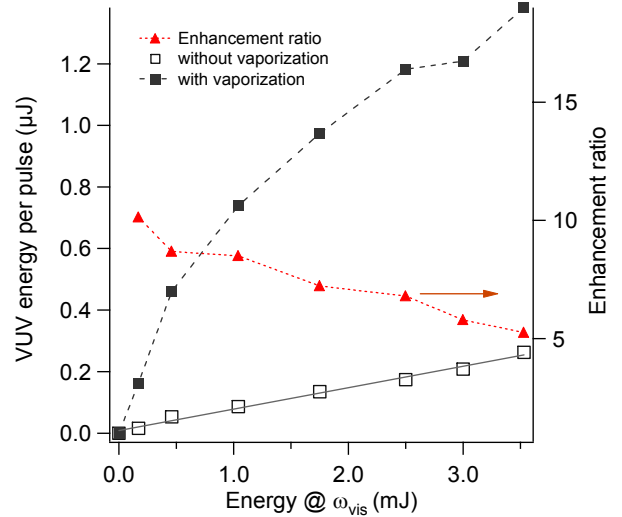

**Figure 5 :** *left axis*: VUV energy per pulse obtained for a fixed energy of UV beam (2.4 mJ). *Open squares*: without vaporization; the line is a linear fit showing that no saturation occurs. *Full squares*: with vaporization (the line is a guide for the eye). *Right axis*: enhancement ratio (VUV energy with vaporization / energy without vaporization). The excimer energy is 2 mJ (0.2 J/cm² on Hg surface).

We found that the enhancement factor noticeably increased when either the UV or the visible beam intensity was decreased, as shown in Figure 4 for a fixed energy of the red beam (3.4 mJ), and in Figure 5 for a fixed energy of the UV beam (2.4 mJ). As seen in Figure 4, the enhancement attained values >150 for UV energies below 0.3 mJ (power densities < 0.5 GW/cm²). Higher values were obtained for lower fluences but the corresponding VUV signal obtained without vaporization was not retrievable from noise in this case. The same saturation behavior was observed as a function of the visible energy (Figure 5), while keeping UV at maximum power. The VUV signal obtained when the excimer laser was turned off showed the expected dependance over the mixing fields in absence of saturation, that is linear with the energy at $\omega_{vis}$ and quadratic with the energy at $\omega_{UV}$ (cf. figures 4 et 5). The two-photon transition clearly plays a key role in the observed saturation, as established by the fact that once the UV energy has been fixed to a high value (figure 5), the enhancement factor is modest and much less dependant on the energy at $\omega_{vis}$ than it was dependant on the UV energy (figure 4). We checked that at a lower UV energy of 1.2 mJ, the enhancement was much higher (> 50) for a 0.2-mJ red beam.

Saturation behavior in two-photon resonant mixing processes is typical of the interaction of intense fields with dense gaseous media, and has been discussed by many authors [29, 30]. Two-photon absorption plays a key role here since the two-photon absorption cross section for the $6\,^1S_0 \rightarrow 7\,^1S_0$ transition is especially high ($2.6\,10^{-6}\,s^{-1}.W^{-2}.cm^4$ [31]). With UV intensities in the GW/cm² range (here, 3.5 GW/cm² for $\omega_{UV}$), the transition can be theoretically saturated. This yields to a depletion of the ground state, as well as an increase of the number of $Hg^+$ ions due to three-photon photoionization. These effects reduce the number of useful photons for 4-wave mixing, and strongly modify phase-matching conditions.

The detailed modelling of our experimental configuration is particularly intricate. A comprehensive study of four-wave mixing in mercury, including saturation issues, has been published by Smith *et al.* [27]. However the model taken from this paper is not relevant in our case, since it was restricted to low intensities of mixing fields (~MW/cm²), that is a few orders of magnitude lower than in this work. Thus, competing nonlinear effects (amplified spontaneous emission from 7s to 6p levels, hyper-Raman scattering, fifth-order nonlinear effects, etc.) cannot be safely ignored. Moreover, Smith *et al.* distinguished two different mechanisms for saturation, one for exact resonance and the other for off-resonance situations: this distinction is not appropriate here since the spectral width of our laser is much higher than the Doppler-broadened profile of one isotope. A complete description requires further investigation and is beyond the scope of this paper.

### 5. Broadening of the two-photon resonance

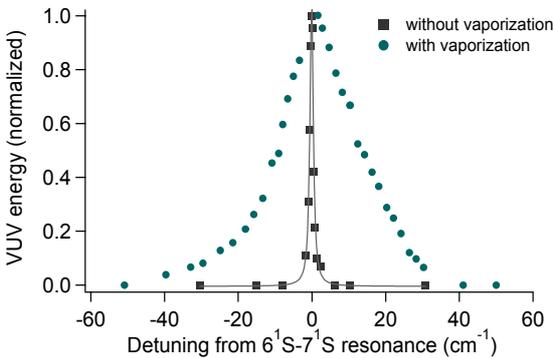

**Figure 6 :** VUV yield (normalized) versus the frequency detuning $\omega_{6s-7s} - 2\omega_{UV}$ (the detuning is indicated with respect to the central two-photon 6s-7s transition, i.e. 63928.243 cm$^{-1}$.) The excimer fluence on Hg was 0.2 J/cm².

In order to get more experimental information about the role of the 2-photon transition on the observed saturation, we tuned $\omega_{VIS}$ so that the UV frequency sweeps across the $6\,^1S_0 \rightarrow 7\,^1S_0$ resonance (see Figure 6). Without the vaporizing laser, a lorentzian fit indicates a FWHM of 0.7 cm$^{-1}$; all the widths in the following are given with respect to the frequency of the $6\,^1S_0$-$7\,^1S_0$ transition at 63928.243 cm$^{-1}$. This value is consistent with the finite linewidth of the laser — 0.4 cm$^{-1}$ when expressed relative to the two-photon transition energy— and with the fact that naturally-occuring mercury is a mix of 7 different isotopes spreading over ~0.5 cm$^{-1}$ (the Doppler width is only 0.06 cm$^{-1}$). In contrast, a large broadening was noticed when the vaporizing laser was turned on. When the UV energy was 2 mJ, a FWHM of 26 cm$^{-1}$ was measured. This means that very large enhancements (>100) can be obtained when the UV frequency is tuned off resonance; when the laser dye wavelength at $\omega_{vis}$ is detuned by more than 0.1 nm (that is ~6 cm$^{-1}$ detuning from exact 2-photon resonance) a strong VUV signal is detectable only in presence of vaporization. The broadening turned out to be very dependant on the UV intensity, as seen in Figure 7.

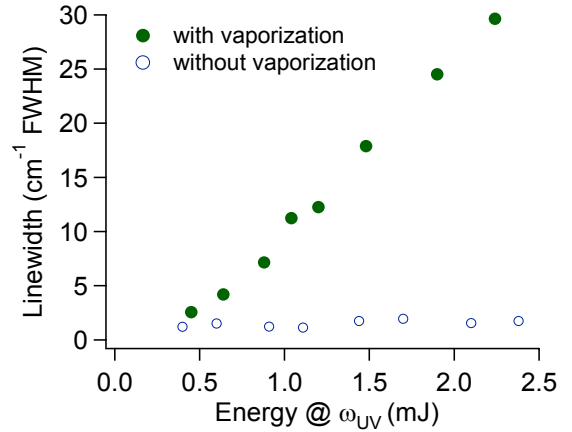

**Figure 7 :** Linewidth (at half maximum) of VUV signal around the 2-photon resonance vs. energy of the UV beam at 313 nm, when the nonlinear interaction takes place in the ablation plume (*full circles*), and in the homogeneous vapor (*open circles*). The excimer fluence on Hg was 0.2 J/cm².

The linewidth increased linearly with the UV intensity when the vaporizing laser was turned on, whereas no broadening was measurable in the room-temperature homogeneous vapor.

We now discuss the origin of this intensity-dependant broadening. From Figure 7, some effects can first be ruled out (or assigned to a minor role): Doppler broadening cannot be responsible for an intensity-dependant broadening, except if the UV beam was heating the plasma: this is very unlikely since, the Doppler width scaling as the square root of temperature, a 30 cm$^{-1}$ broadening would mean unrealistic temperatures (~10$^8$ K) inside the plume. Pressure broadening can be ruled out for the same reasons. Power broadening seems also unlikely, whereas it is at first sight suggested by the intensity dependance: indeed, for an ideal homogeneously broadened two-photon transition, the linewidth is given by [32]:

$$\Delta \nu = \Delta \nu_0 \left(1 + 2\tau \sigma^{(2)} I^2\right)^{1/2} \quad (1)$$

where $\Delta\upsilon_0$ is the homogeneously broadened linewidth at zero intensity, $\tau$ the effective lifetime of excited states, $\sigma^{(2)}$ the two-photon absorption cross section and $I$ the UV intensity (W.cm$^{-2}$). However, given that collisions and higher pressure inside the plasma would tend to lower the effective lifetime of the excited state rather than increase it, we should expect a lower slope in Figure 7 in the ablation plume than in the homogeneous vapor, which is the opposite of the experimental findings.

It is likely that this large broadening is a result of several nontrivial processes. Stark broadening cannot be excluded, as it is a well-known cause of line broadening in laser-created plasmas [33], however the intensity-dependance could only be the result of electrons and ions created upon UV photoionization. It has been reported that a broadening of the 7s level can result from competing effects that deplete this level, such as resonant photoionization or Amplified Spontaneous Emission (ASE) towards the 6 $^1$P level [34]. Furthermore, the large VUV signal obtained far from resonance may be an indication that off-resonant mixing is favored by some change of the nonlinear phase mismatch, caused by a two-photon-resonantly enhanced Kerr effect. We believe that the observed broadening is a combination of these effects altogether; a fine modelling would allow estimating the relative contribution of each effect.

At last, it is important to notice that phase-matching conditions in a highly inhomogeneous ablation plume are different than in a standard homogeneous vapor. In this latter case, it is well known that the wavevector mismatch $\Delta k = k_{VUV} - 2k_{UV} - k_{vis}$ has to be negative in a tight focussing geometry [35]. Remarkably, it has been noted by several authors [21, 28, 30] that the strong signal at 125.14 nm is observed in a region of positive dispersion for mercury, which has not been clearly elucidated up to now. Further improvements in the source efficiency will require to better understand how phase matching governs the VUV yield, in relation with the density spatial distribution of mercury.

## 6. Conclusion

In conclusion, we have shown an original way to increase the yield of a VUV coherent source at 125 nm, obtained from one single dye laser at 625.7 nm, without heating the overall Hg cell. An ArF excimer laser at 193 nm was sent to a Hg surface 0.7 µs before the nonlinear interaction occurred. When the UV beam (1.6 mJ, that is 2.3 GW/cm$^2$ at focus) was tuned to the 6 $^1$S$_0 \rightarrow$ 7 $^1$S$_0$ two-photon resonance ($\lambda$=312.85 nm), an enhancement of the VUV power of ~6 was measured with respect to the power measured in a room-temperature Hg vapor. A study of the enhancement versus the excimer pulse energy has put into evidence the existence of a threshold (30 mJ/cm²) and a saturation plateau (> 100 mJ/cm²). We noticed that the enhancement could reach very high values (> 150) when either the UV or the red beams had lower intensities, or alternatively when the frequency of the dye laser was tuned slightly off-resonance of the 2-photon transition. This illustrated the existence of very strong saturation effects, connected to the 6 $^1$S$_0 \rightarrow$ 7 $^1$S$_0$ transition. Two-photon absorption and photoionization play key roles in reducing the efficiency at high powers. However we have shown that a fine modelling is intricate since no competing effect can *a priori* be ignored. A further evidence of the role of the two-photon transition was brought by the measurement of the VUV signal versus the frequency of the two-photon resonant UV beam; a considerable broadening has been measured (up to 30 cm$^{-1}$ FWHM with vaporization against 0.7 cm$^{-1}$ in the room-temperature vapor), which varied linearly with the UV power. Whereas this behavior suggested power broadening, we showed that this was unlikely in this situation, and discussed several other mechanisms. This work opens some perspectives for the achievement of high-efficiency and compact VUV sources: by taking into account phase matching effects more in detail, the source has the potential to be further optimized (e.g. by varying the size of the UV and visible beams inside the plasma, tailoiring the shape of the plume, etc.) Additionally, the use of coherent preparation of states (such as SCRAP [20]) may be envisaged as a way to overcome saturation issues. At last, one can take advantage of the large broadening of the two-photon resonance discussed in this paper to release the constraint on the narrow linewidth of the tunable dye laser: the source could be made significantly more compact using an all-solid-state compact OPO (optical parametric oscillator).